\documentstyle[aps,array,floats,twocolumn,epsf]{revtex}
\begin{document}
\def\be{\begin{equation}}
\def\ee{\end{equation}}
\def\bea{\begin{eqnarray}}
\def\eea{\end{eqnarray}}
\def\E{{\rm e}}
\def\bearst{\begin{eqnarray*}}
\def\eearst{\end{eqnarray*}}
\def\peleven{\parbox{9cm}}
\def\peffec{\peight{\bearst\eearst}\hfill\peleven}
\def\pspace{\peight{\bearst\eearst}\hfill}
\def\ptwelve{\parbox{12cm}}
\def\peight{\parbox{8mm}}
\twocolumn[\hsize\textwidth\columnwidth\hsize\csname@twocolumnfalse\endcsname 
\title
{Theoretical Model for Kramers-Moyal's Description of Turbulence 
Cascade} \author
{Jahanshah Davoudi$^{a,c}$ and 
M. Reza Rahimi Tabar$^{a,b}$}
\address
{\it $^a$ Institute for Studies in Theoretical Physics and 
Mathematics
Tehran P.O.Box: 19395-5531, Iran,\\
$^b$ Dept. of Physics, Iran  University of Science and Technology,
Narmak, Tehran 16844, Iran. \\
$^c$
Dept. of Physics, Sharif University of Technology,
Tehran P.O.Box:11365-9161, Iran}

\maketitle

\begin{abstract}
We derive the Kramers-Moyal equation for the conditional probability density  
of velocity increments from the theoretical model 
recently proposed by V.Yakhot \cite{1} 
in the limit of high Reynolds number limit.
We show that the higher order ($n \geq 3 $ ) Kramers -- Moyal  coefficients 
tends to zero and the velocity increments are evolved by the Fokker-- Planck 
operator.
Our result is compatible with the phenomenological
descriptions by R.Friedrich and J.Peinke \cite{2}, developed for explaining 
the experiments recently done by J. Pienke. et al.\cite{3}. 

\end{abstract}
\vskip 0.2in
]

The problem of scaling behavior of longitudinal velocity difference 
$U=|u(x_1) - u(x_2)|$
in turbulence and the probability density function of $U$ i.e $P(U)$,
attracts a great deal of attention [4-10].
Statistical theory of turbulence has been put
forward by Kolmogorov [11], and further developed by others [12-15].
The approach is to model turbulence using stochastic
partial differential equations.  
Kolmogorov conjectured that the scaling 
exponents are universal, independent of the statistics of large--scale
fluctuations and the mechanism of the viscous damping, when
the Reynolds number is sufficiently large. 
However, recently 
it has been found that there is a relation between the 
probability distribution function (PDF) of velocity and those of the  
external force (see [1] for more detail).
In this direction, Polyakov [5] has 
recently offered a field theoretic method to derive the 
probability distribution or density of states in (1+1)-dimensions 
in the problem of randomly driven Burgers equation [16-17]. 
In one dimension, turbulence without pressure
is described by Burgers equation (see also [18] concerning the
relation between
Burgers equation and KPZ--equation). 
In the limit of high Reynolds number, using 
the operator product expansion (OPE), Polyakov reduces the problem of 
computation of correlation functions in the inertial subrange, 
to the solution of a certain partial 
differential equation [19-20].  Yakhot recently [1,23] generalize the
Polyakov approach in three-dimensions and find a closed differential
equation for the two-point generating function of the $"longitudinal"$
velocity difference in the strong turbulence (see also [21] 
about closed equation
for PDF of velocity difference for two and three-dimensional turbulence
without pressure). 
On the other hands, recently [2] from detailed analysis of experimental data of a turbulent free jet,
R.Friedrich and J.Pienke have been able to obtain a phenomenological description 
of the statistical properties of a turbulent cascade using a Fokker-Planck 
equation. In other words they have seen that the conditional probability density
of velocity increments satisfy the Chapman-Kolmogorov 
equation. Mathematically this is a necessary
condition for the velocity increments to be a Markovian process in terms 
of length scales.
By fitting the observational data they have succeeded to find the different
Kramers-Moyal(K.M) coefficients and they find that the approximations
of the third and fourth order coefficients tend to zero whereas the first and 
second coefficients have well defined limits. Then giving address to the 
implications dictated by  [22]
theorem they have got a Fokker-Planck evolution operator. As an evolution 
equation for the probability density function of velocity increments, the 
Fokker-Planck equation has been used to give the information on changing shape
of the distribution as a function of the length scale. By this strategy the
information on the observed intermittency of the turbulent cascade is 
verified. In their
description and based on simplified assumptions on the drift and diffusion coefficients
they have considered two possible scenarios in order to indicate that both the
Kolmogorov 41 and 62 scalings are recovered as possible behaviors in
their phenomenological theory.\\
In this paper we derive the Kramers--Moyal equation from Navier--Stokes
equation and show that how the higher order ($n \geq 3$) Kramers--Moyal 
coefficients tend to zero in the high Reynolds number limit.
Therefore we find the Fokker--Planck equation from first principles.
We show that the breakdown of the Galilean invariance
is responsible for scale dependence of the Kramers--Moyal coefficients.
Finally using the path-integral expression for the PDF we show that how small
scale statistics affected by PDF`s in the large scale and confirm the Landau`s
remark that the large-scale fluctuations of turbulence production in the 
integral range can invalidate the Kolmogorov theory [12-13].

Our starting point is the Navier--Stokes equations:

\be
{\bf v}_t + ({\bf v} \cdot \nabla ) {\bf v} = \nu \nabla^2 {\bf v} - 
\frac {\nabla p}{\rho} + {\bf f}({\bf x},t), \hskip .5cm \nabla \cdot \bf{v}=0
\ee

for the Eulerian velocity $ {\bf v}({\bf x},t)$ and the pressure
$p$ with viscosity $\nu$, in N--dimensions.
The force $ {\bf f}({\bf x},t)$ is the external stirring force, which
injects energy into the system on a length scale $L$.
More specifically one can take, for instance a Gaussian distributed  
random force, which is identified by its two moments:

\be
< f_\mu ({\bf{ x}},t)  f_\nu ({\bf x^{'}},t^{'})> =  
k(0) \delta (t-t^{'}) k_{\mu \nu}({\bf {x}- { x^{'}}})
\ee
and $< f_\mu ({\bf{ x}},t)> = 0 $,     
 where $\mu, \nu = x_1, x_2, \cdots ,x_N$. The correlation function 
$k_{\mu \nu}(r)$ is normalized to unity at the origin and decays 
rapidly enough where $r$ becomes larger or equal to integral scale $L$.

The force free N-S equation is invariant under space--time translation,
parity and scaling transformation. Also it is invariant under Galilean
transformation, $x \rightarrow x + V t$ and $v \rightarrow v + V$, where
$V$ is the constant velocity of the moving frame. 
Both boundary conditions and 
forcing can violate some or all of symmetries of force free N-S equation.
However it is, usually assumed that in the high Reynolds number flow 
all symmetries of the N-S equation are restored in the limit 
$r \rightarrow 0$ and $ r >> \eta$, where $\eta$ is the dissipation 
scale where the viscous effects become important. This means that in this 
limit the root--mean square velocity fluctuations $u_{rms}= \sqrt{<v^2>}$
which is not invariant under the constant shift $V$, cannot enter the relations
describing moments of velocity difference. Therefore the effective
equations for the inertial--range velocity correlation functions must have 
the symmetries of the original N-S equations. For many years this 
assumption was the basis of turbulence theories. But based on the recent
understanding of turbulence, some of the constraints on 
the allowed turbulence theories can be relaxed [1].
Polyakov's theory of the large--scale random force driven Burgers
turbulence [5] was based on the assumption that weak small -- scale
velocity difference fluctuations (i.e. $ | v(x+r) - v(x) | << u_{rms} $
and $ r << L$),  where $L$ is the integral scale of system, obey 
$G$--invariant dynamic equation, meaning that the integral scale and the
single--point $u_{rms}$ induced by random--forcing cannot enter the
resulting expression for the 
probability density. According to [1] it has been shown that how 
the $ u_{rms}$ enters the equation for the PDF and therefore breaks 
the G-invariance in the limited Polyakov`s sense.
We are interested in the scaling  of the 
longitudinal structure function $S_q = < (u(x+r) - u(x))^q> = < U^q >$,
where $u(x)$ is the $x$-component of the three-dimensional velocity field
and $r$ is the displacement in the direction of the $x$-axis and the
probability density $P(U,r)$ for homogeneous and isotropic turbulence. 
Let us define the generating function 
$\hat Z$ for longitudinal structure function $\hat Z = <e ^{\lambda U}>$.
According to [1] in the spherical coordinates the advective term in 
eq.(1) involve the terms 
$O(\frac{\partial^2 \hat Z}{\partial \lambda \partial r})$
, $O(\frac{\partial \hat Z}{ r \partial \lambda} )$,  
$O(\frac{\partial \hat Z}{ \lambda \partial r}) $,
$O(\frac{\hat Z}{ \lambda  r})$ [21]. 
It is noted that the advection contributions are accurately accounted 
for in equation of $\hat Z$, but it is not closed due to the
dissipation and pressure terms. Using the Polyakov`s OPE approach Yakhot
has shown the 
the dissipation term can be treated easily while the pressure term has an 
additional difficulty. 
The pressure contribution leads to effective 
energy redistribution between components of velocity field and has 
non-trivial effect in the dynamics of N-S equation. 
 Proceeding to find a closed equation for 
the generating function of Longitudinal velocity difference, $\hat Z$,
the dissipation and pressure terms in eq.(1) give contributions and 
the longitudinal part of the dissipation term renormalizes the coefficient in front of 
the $O (\frac {1}{\lambda})$ in equation for $\hat Z$, [1]. Also it 
generates a term with order of $O(u)$ which can be written in terms of $\hat Z$ as 
$\lambda \frac {\partial \hat Z}{\partial \lambda}$. Taking into account 
all the possible terms and using the symmetry of the PDF i.e $P(U,r) = 
P(-U,-r)$, 
the following closed equation for $\hat Z$ can be found,[1],
\be  \frac{\partial^2 \hat Z}{\partial \lambda \partial r} - \frac {B_0}
{\lambda} \frac{\partial \hat Z}{ \partial r} 
= \frac {A}{r} \frac{\partial \hat Z}{ \partial \lambda} - C \lambda   
\frac{\partial \hat Z}{ \partial \lambda} + 3 r^2 \lambda^2 \hat Z 
\ee
 where the parameter $A$, $B$ and $C$ to be determined from the theory.
Also we suppose that $k_{\mu \nu}$ has the structure $
k_{\mu \nu} ({\bf r_{i,j}})
 = k(0) [1- 
\frac {| {\bf r_{i,j}}
|^2}{2 L^2} \delta_{\mu,\nu} - \frac {({\bf r_{i,j}})_\mu 
( {\bf r_{i,j}})_\nu}{L^2} ]
$
with $k(0)=1$ and ${\bf r_{i,j}}={\bf x_i} - {\bf x_j}$. 
The Gaussian assumption for $"single-point"$ probability density fixes the 
 value of coefficient $C=\frac{u_{rms}}{L}$ and the $C$-term corresponds
 to the breakdown of G--invariance in the limited Polyakov's sense [5].
 The $A$-term is responsible for interaction of transverse components of 
 velocity field with the longitudinal component and produce an 
 effective source and friction for the longitudinal correlation.

In the limit $r \rightarrow 0$ the equation for the probability density
is derived from eq.(3) as,
\be  -\frac{\partial}{\partial U} U \frac{\partial P}{ \partial r} -  B_0
\frac{\partial P}{ \partial r} 
= - \frac {A}{r} \frac{\partial} { \partial U} U P + \frac {u_{rms}}{L}
\frac{\partial^2}{ \partial U^2} U P 
\ee

Using the exact results $S_3 = - \frac{4}{5} \epsilon r $ in the small scale,
( $\epsilon$ is the mean energy dissipation rate) one finds 
$A= \frac {3+B}{3}$, where $B=-B_0 >0$ [1]. 
It is easy to see that the eq.(4) can be written as
$\partial_r P=(-\partial_{U} U-B_0)^{-1}[-(A/r)\partial_{U} U + (u_{rms}/L)
\partial_{U}^{2} U]P$ and so its solution can obviously be written as a 
scalar--ordered exponential [22], $P(U,r)={\cal 
T}(e_{+}^{\int_{r_0}^{r} dr' L_{KM}(U,r')} P(U,r_0))$
, where $L_{KM}$ can be obtained formally by computing the inverse operator.
Using the properties of scalar--ordered exponentials the conditional 
probability density will satisfy the Chapman-Kolmogorov
equation. Equivalently we derive that the probability density
and as a result the conditional probability density of velocity increments 
satisfy a K.M. evolution equation:
\be
-\frac{\partial{P}}{\partial{r}}=\sum_{n=1}^{\infty} {(-1)}^{n}\frac{\partial^{n}}{\partial{U}^{n}}(D^{(n)}(r,U)P)
\ee
Where $D^{(n)}(r,U)=\frac{\alpha_{n}}{r} U^{n}+\beta_{n} U^{n-1}$.
We have found that the coefficients $\alpha_{n}$ and $\beta_{n}$ depend on $A$ and
$B$, $u_{rms}$ and inertial length scale $L$ which are given by the recursion
relations
We scale the velocities as 
$\tilde{U}=\frac{U}{(\frac{r}{L})^{1/3}}$
and introduce a logarithmic length scale $\lambda=\ln(\frac{L}{r})$ which varies
from zero to infinity as $r$ decreases from $L$ to $\eta$.
Thus the form of $\tilde{D^{(1)}}(\tilde{U},r)$ and $\tilde{D^{(2)}}(\tilde{U},r)$ in the equivalent
description would be $\tilde{D}^{(1)}(\tilde{U},r)=-(\frac{A}{1+B})\tilde{U}
$ and
$\tilde{D^{(2)}}(\tilde{U},r)=(\frac{A}{(2+B)(1+B)})\tilde{U}^2-(\frac{r}{L})^{2/3}u_{rms}(\frac{(1)}{(2+B)})\tilde{U}$.

The drift and diffusion coefficients for various scales $\lambda$, determined
in the theory of Yakhot,
show the same functional form as the calculated coefficients 
from experimental data [2,3].

In comparison with the phenomenological theory of Friedrich and Pienke we are
able to construct a K.M. equation for velocity increments that is analytically
derived from Yakhot theory which is based on just
general underlying symmetries and OPE conjecture. Furthermore this viewpoint
on the equation (4) gives the expressions for scale dependence of the 
coefficients in the K.M. equation. 
The important result
is that scale dependent K.M.
coefficients are proportional to $u_{rms}$ which gives a probable relationship
between breakdown of G-invariance and scale dependence of the K.M. coefficients
in the equivalent theory.
The two unknown parameters $A$ and $B$ in the theory is reduced to
one by fitting the $\xi_{3}=1$, so all the scaling exponents and
$D^{(n)}$'s are described by one parameter, $B$. 
Considering the results in [1,2] on which the value of $B$ are obtained,
we have used the value 
$B\cong 20$ and have calculated the numerical values of KM coefficients.
Ratios of the first three coefficients $\alpha_{n}$ and 
$\beta_{n}$ are $\alpha_3 / \alpha_2=0.04, \alpha_4 / \alpha_2=0.001
, \beta_3 / \beta_2=0.04, \beta_4 / \beta_2=0.001$.
From the comparison of numerical values of higher order coefficients we find that 
the series can be cut safely after the second term and a good 
approximation for  
evolution operator of velocity increments is a Fokker-Planck operator.
According to [1] the value of parameter $B\cong20$ is calculated 
numerically in the limit of infinite Reynolds numbers. 
Using this value for calculation of the numerical values 
of $\tilde{D_{1}}$ and $\tilde{D_{2}}$ we find that the contribution 
of scale dependent terms are essentially negligible. 
This correspondence was referred to in [1] about the behavior of 
diffusion coefficients for the limit of infinite Reynolds numbers.
As it is well-known the Fokker-Planck description of Probability measure 
is equivalent with the Langevin description written as [22],
$\frac{\partial{\tilde{U}}}{\partial{\lambda}}=\tilde{D}^{(1)}(\tilde{U},\lambda)+\sqrt{\tilde{D}^{(2)}(\tilde{U},\lambda)}\eta(\lambda)$,
, where $\eta(\lambda)$ is a white noise and the diffusion term acts as a 
multiplicative
noise. By considering the Ito prescription and using the path-integral
representation of the Fokker-Planck
equation we can give an expression for all the possible paths in the
configuration space of velocity differences and thus demonstrate the
change of the measure under the change of scale,i.e.
\be
P(\tilde{U}_{2},\lambda_{2}|\tilde{U}_{1},\lambda_{1})=\int {\cal D}[\tilde{U}]
e^{-\int_{\lambda_{1}}^{\lambda_{2}} d\lambda \frac{ {(\frac{\partial{\tilde{U}}}{\partial{}\lambda}-{\tilde{D}}^{1}(\tilde{U},\lambda))}^2}{4{\tilde{D}^{2}(\tilde{U},\lambda)}}}
\ee
When calculating, the measure of path integral is meaningful when some
form of discretization is chosen [22], but we have written it in a formal
way.
Using the forms of $\tilde{D}^{1}$ and $\tilde{D}^{2}$ and approximating them
with scale independent ones in the infinite Reynolds number limit, one can easily
see that the  transition functional can be written in terms of 
$\ln{\tilde{U}}$.
It is an easy way to see how the large scale $\lambda\rightarrow{0}$ Gaussian probability density can
change its shape when going to small scales $\lambda\rightarrow{\infty}$ and
consequently give rise to intermittent behavior.
Instead of working with the probability Functional of velocity increments
the formal solution of Fokker-Planck equation as a scalar--ordered 
exponential [25],
can be converted to an integral representation for the probability measure of velocity
increments when the $\tilde{D}^{1}\cong-\alpha_{1}(\lambda)\tilde{U}$
and $\tilde{D}^{2}\cong\alpha_{2}(\lambda)\tilde{U}^{2}$ ,i.e
\be
P(\tilde{U},\lambda)=\frac{e^{\gamma_{0}(\lambda)}}{\sqrt{4\pi\gamma(\lambda)}}\int_{-\infty}^{+\infty} e^{-\frac{s^{2}}{4\gamma(\lambda)}}\phi(\tilde{U}e^{\gamma_{1}(\lambda)-s}) ds
\ee
where, $\gamma_{0}(\lambda)=\int_{0}^{\lambda}(-\alpha_{1}(\lambda')+2\alpha_{2}(\lambda'))d\lambda'$ and 
$\gamma_{1}(\lambda)=\int_{0}^{\lambda}(-\alpha_{1}(\lambda')+3\alpha_{2}(\lambda'))d\lambda'$ and $\gamma(\lambda)=\int_{0}^{\lambda}\alpha_{2}(\lambda') d\lambda'$
and $\phi(\tilde{U})$ is the Probability measure in the integral length scales $(\lambda\rightarrow 0)$ .  
We consider the Gaussian distribution, $\phi(\tilde{U})\cong e^{-m\tilde{U}^{2}}$ in the 
integral scale which is a reasonable 
choice ( experimental data shows that up to third moments 
the PDF in the integral scale are consistent with Gaussian distribution [1]) 
and derive the dependence of the variance of the probability density 
on the scales in the limit 
when the original distribution satisfies the condition $m\ll 1$. The result shows an exponential dependence
like $m\rightarrow m e^{2\zeta}$ where $\zeta=3\alpha_{2}-\alpha_{1}$.
The consistent picture with the shape change of probability measure under the
scale is that when $\lambda$ grows, the width decreases and vice versa,
which is reported in previous works as a simulation and experimental results.
Moreover we should emphasize that the shape change is somehow complex which
gives some corrections in order $O(m^2\tilde{U}^{4})$ even in this simplifying limit,i.e.
$m\ll 1$. Starting with a Gaussian measure at integral scales  and using the
calculated scale independent Fokker-Planck coefficients, we have numerically
calculated the PDF's for fully developed turbulence and Burgers turbulence 
in different length scales which their plots in Fig.[1] and Fig.[2] 
are completely compatible with experimental and simulation results [1,2,4].
The extreme case of Burgers problem (i.e.$B\cong 0$) shows the ever 
localizing
behavior as if in the limit of $\lambda\rightarrow {\infty}$ goes to a
Dirac delta
function which again is consistent with our knowledge about Burgers problem [1,4].
Clearly the eqs. (4) and (5) give the same result for multifractal 
exponent of structure function 
,i.e. $S_{n}(r)\cong A_{n}r^{\xi_{n}}$ is derived to be 
$\xi_{n}=\frac{(3+B)n}{3(n+B)}$ [1].\\
In summary We have constructed a theoretical bridge between two recent theories
involving the statistics of longitudinal velocity increment fluctuations in fully
developed Turbulence. On the basis of the recent theory proposed by V. Yakhot 
we showed that the probability density of longitudinal velocity components
satisfy a Kramers-Moyal equation which encodes the Markovian property of these
fluctuations in a necessary way. We are able to give the exact form of Kramers-Moyal 
coefficients in terms of a basic parameter in Yakhot theory $B$. The qualitative
behavior of drift and diffusion terms are consistent with the experimental 
outcomes [2]. 
As the most prominent result of our work, we could find the form of path 
probability
functional of the velocity increments in scale which naturally encodes the
scale dependence of probability density. 
This gives a clear 
picture about the intermittent nature in fully developed Turbulence.

We should emphasize that the derivation of KM equation is not restricted 
to the Polyakov's specific approach. One can show that 
similar results could
be obtained by the conditional averaging methods [24-25]. Clearly 
analytic form of the K.M.
coefficients $D^{(n)}$ can be estimated numerically but analytic 
derivation is not possible [26]. 
Our work might be generalized to give a theoretical basis for the Markovian 
fluctuations of the moments of height difference in the surface growth problems 
like KPZ [18,27] and we believe that it would be possible to derive 
the Kramers-Moyal description for the statistics of energy
dissipation[28].\\
\vskip 0.08cm
{\bf Acknowledgement}: We would like to thank from A. Aghamohamadi, B. 
Davoudi, R. Ejtehadi, M. Khorrami, A. Langari and S. Rouhani for helpful
discussions and useful comments.

\vspace{40cm} 

\begin{center}
{\bf FIGURE CAPTIONS}
\end{center}

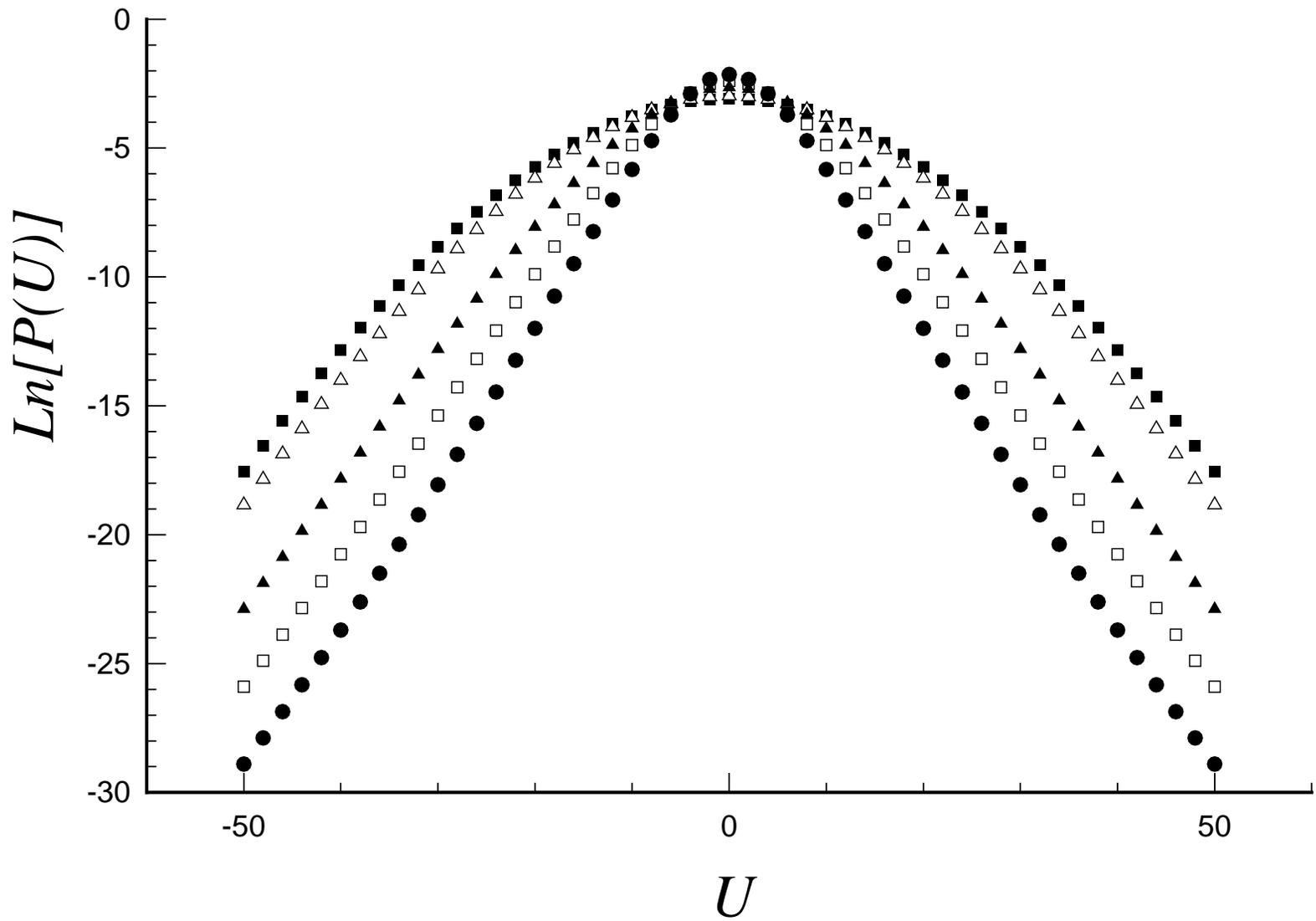
Figure 1. {Schematic view of the logarithm of PDF in terms of
different length scales. These graphs are numerically obtained from the 
integral representation of PDF at the Fokker-Planck approximation.
The curves correspond with the scales $L/r=1.5,2,5,10,20$.}\\ 
\\

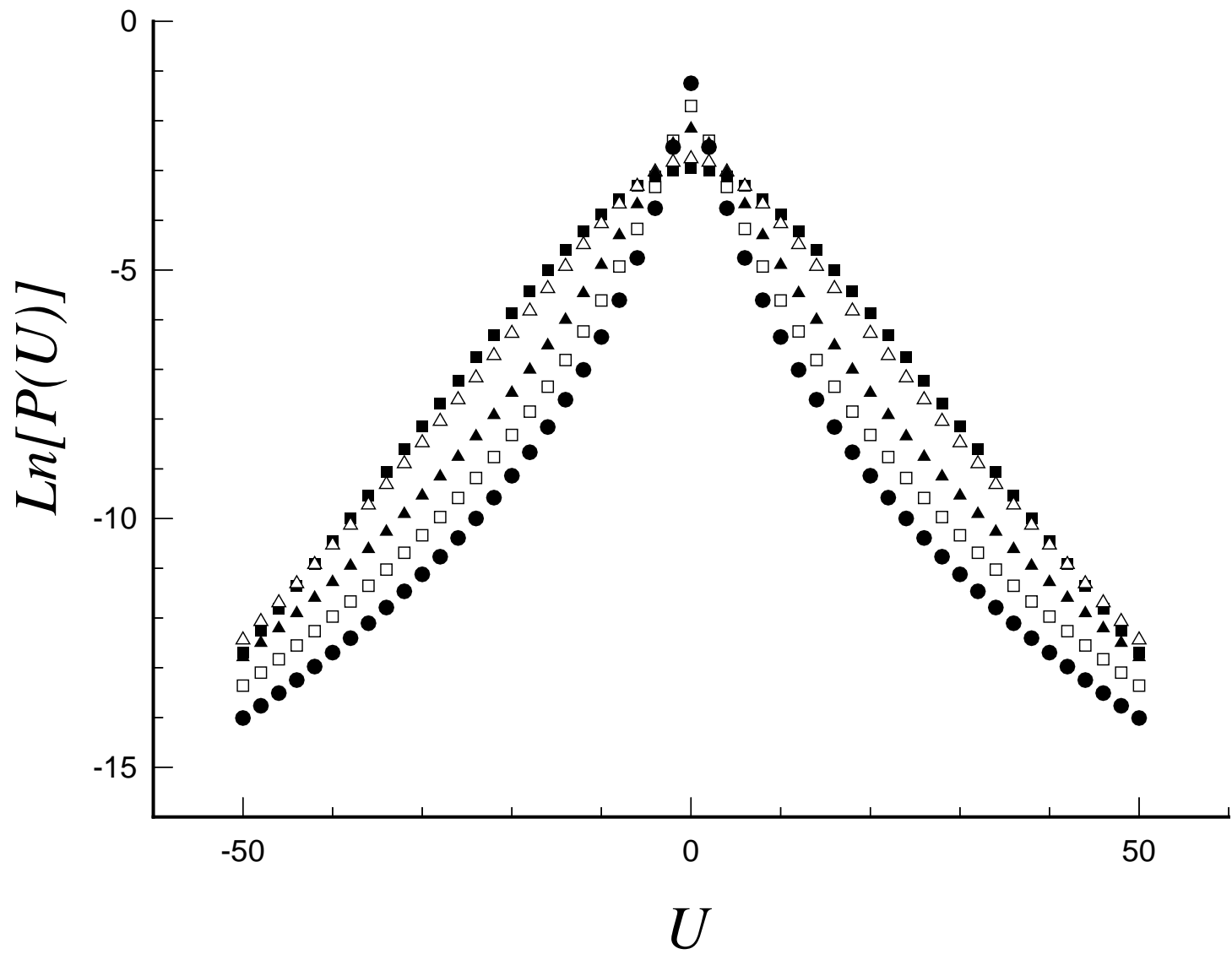
Figure 2. {Schematic view of the logarithm of PDF 
in the Burgers turbulence ($B\cong0$), in terms of
different length scales.These graphs are numerically obtained from the 
integral representation of PDF at the Fokker-Planck approximation.
The scales are $L/r=1.5,2,5,10,20$.}

\end{document}